\documentclass[final]{aa}
\usepackage[final]{graphicx}

\newcommand {\be} {\begin{equation}}
\newcommand {\ee} {\end{equation}}
\newcommand {\bea} {\begin{eqnarray}}
\newcommand {\eea} {\end{eqnarray}}

\begin{document}

\title{Numerical Simulations of Radiation from Blazar Jets}

\author{R.~Moderski \and M.~Sikora \and M.~B{\l }a\.zejowski}

\institute{Nicolaus Copernicus Astronomical Center,
		Bartycka 18, 00716 Warsaw\\
		\email{moderski@camk.edu.pl}}

\date{Received ; accepted -- }

\abstract{ 
  We present a description of our numerical code $BLAZAR$.
  This code calculates spectra and light curves of blazars during
  outbursts. The code is based on a model in which the non-thermal
  flares in blazars are produced in thin shells propagating down a
  conical jet with relativistic velocities. Such shells may represent
  layers of a shocked plasma, enclosed between the forward and reverse
  fronts of an internal shock. In the model adopted by us, the
  production of non-thermal radiation is assumed to be dominated by
  electrons and positrons which are accelerated directly, rather then
  injected by pair cascades.  The code includes synchrotron emission
  and inverse-Compton process as the radiation mechanisms.  Both
  synchrotron photons and external photons are included as the seed
  photons for Comptonization. At the present stage, the code is
  limited to treat the inverse Compton process only within the Thomson
  limit and is specialized to model radiation production in the flat
  spectrum radio quasars.  As an example, we present the results of
  modeling an outburst in 3C 279 -- the most extensively monitored
  $\gamma$-ray - bright quasar.
   }

\titlerunning{Radiation from Blazar Jets}
\authorrunning{Moderski et al.}
\maketitle
\section{Introduction}
One of the greatest achievements of the recently retired Compton
Gamma-Ray Observatory (CGRO) is the discovery of gamma ray emission
from a subclass of Active Galactic Nuclei (AGNs) known as blazars.
More than $50$ such sources were detected at energies above $100$MeV
(Mukherjee et al.~\cite{muk97}).  Energy constraints and $\gamma$-ray
absorption arguments require that radiation emission in blazars must
be beamed (Mattox et al.~\cite{mat93}).  Indeed, it is now widely
believed that the entire electromagnetic spectrum observed in these
objects is dominated by non-thermal radiation produced in a jet
pointing close to the line of sight (Dondi \& Ghisellini~\cite{dg95}).
Although roughly divided into two classes: flat spectrum radio quasars
(FSRQ) and BL~Lacs objects, all blazars share common characteristics:
large-amplitude, rapid variability; smooth continuum emission in all
observable bands; and high linear polarization.

Spectrum of a blazar consists of two broad components (von Montigny~et
al.~\cite{mon95}).  The low energy component has a peak within
IR-to-X-ray range and is usually attributed to Doppler-boosted
synchrotron radiation.  The high energy component peaks in the MeV-TeV
energy range and is very likely produced by inverse-Compton process.
Both components are highly variable, with time scales ranging from
years to a fraction of a day.  Analysis of $\gamma$-ray light curves
seems to suggest that variability patterns of blazars are a
superposition of short term flares (Magdziarz et al.~\cite{mmm97}).
The flares detected in different spectral bands appear to be
correlated (Macomb et al.~\cite{mac95}; Wagner et al.~\cite{wag95};
Wehrle et al.~\cite{weh98}).  This behavior suggests co-spatial
production of the high and low energy components and indicates that
significant fraction of the jet energy is dissipated in the localized
events at sub-parsec distances from the center.  Such events are
likely to arise in internal shocks (Sikora et al.~\cite{sbr94}; Spada
et al.~\cite{spa01}) or at the sites of the magnetic field
reconnection (Romanova \& Lovelace~\cite{rolo92};
Blackman~\cite{black96}).

Both the shocks and the reconnection sites provide favorable
conditions for efficient acceleration of particles.  Unfortunately,
the present theories of particle acceleration are still not
sufficiently developed to provide quantitative predictions regarding
such issues as what fraction of dissipated energy is used to
accelerate particles, how much power is channeled into relativistic
protons vs. relativistic electrons/positrons, and what maximum energy
can a particle gain.  In particular, it is not possible to deduce
solely on theoretical grounds whether protons are accelerated to
energies sufficient to support -- via inelastic collisions with
photons -- the pair cascades, as is suggested by hadronic models
(Mannhein \& Biermann~\cite{mabi92}), or if they are ``radiatively''
passive returning all energy gained during the dissipative events back
to the flow via adiabatic expansion.  In the latter case, represented
by the leptonic models, radiation production in blazars is totally
dominated by electrons and positrons accelerated directly.  There are
several electron/positron acceleration mechanisms which are likely to
operate in shocks (Levinson~\cite{lev96}; Hoshino et al.~\cite{ho92};
McClements et al.~\cite{mcc97}; Shimada \& Hoshino~\cite{sh00};
Hoshino \& Shimada~\cite{hs02}) and reconnection sites (Larrabee et
al.~\cite{llr02}; Lyutikov~\cite{lyu02}), but such theoretical
considerations cannot as yet predict the spectra of accelerated
particles and their minimum and maximum energies.

Theories of radiation mechanisms are much better developed.  They are
quantitative and for a given energy distribution of the
accelerated/injected particles, geometry and kinematics of the source,
and external radiation field environment, one can make specific
predictions regarding the radiation spectra.  By confronting those
theories with observations, it is possible to verify the model and
determine its parameters.  Using this approach, one can already
exclude some radiation scenarios.  This concerns e.g.\ models with
pair cascades, which predict X-ray spectra that would be too soft as
compared with observations of some FSRQ (Sikora \&
Madejski~\cite{sm01}).  Therefore, for these objects, direct
electron/positron acceleration mechanism is favored.

In leptonic models, both the high-energy and the low-energy components
of the spectrum are produced by the same population of relativistic
electrons.  At low energies, the emission comes from the synchrotron
process, while at higher energies, it is dominated by the inverse
Compton scattering.  There is still some debate about the source of
seed photons for the inverse Compton process.  The most obvious choice
involves synchrotron photons from the low energy component.  Models
based on this assumption are called Self-Synchrotron-Compton (SSC)
models.  Originally proposed by K\"onigl~(\cite{kon81}) to explain
production of X-rays and $\gamma$-rays (already detected by then from
3C~273 by COS-B [Swanenburg et al.~\cite{swa78}]), currently the SSC
models prove to be successful in explaining general features of BL Lac
objects (Ghisellini \& Maraschi~\cite{gm89}; Takahashi et
al.~\cite{tak96}; Kirk et al.~\cite{krm98}; Tavecchio et
al.~\cite{tav98}; Kino et al.~\cite{ktk02}).  However, in quasars -
but also, in some radio selected BL Lac objects, the external
radiation fields may be sufficiently dense to dominate the Compton
cooling of electrons in a jet (Madejski et al.~\cite{mad99}).  Models
exploring this hypothesis are called External-Radiation-Compton (ERC)
models.  Several sources of photons for external radiation field have
been considered: direct disc radiation (Dermer \&
Schlickeiser~\cite{ds93}); diffuse radiation from broad emission line
(BEL) region (Sikora et al.\cite{sbr94}); infra-red
radiation from hot dust (B{\l}a\.zejowski et al.~\cite{bsmm00}); or
jet synchrotron radiation scattered back to the jet by the external
gas (Ghisellini \& Madau~\cite{gm96}).

In this paper we present a code developed to simulate non-thermal
flares produced by thin shells propagating down a conical jet with
relativistic speeds.  Such shells approximate the geometry of
relativistic plasma enclosed between the forward and reverse shock
fronts formed by colliding inhomogeneities in a jet (Sikora et
al.~\cite{sbbm01}; Sikora \& Madejski~\cite{sm02}).  The code is
sufficiently general to treat radiation processes for any radial
distribution of magnetic and external radiation fields, and for any
electron/positron injection function.  It includes both Comptonization
of synchrotron radiation and Comptonization of external radiation, and
takes into account adiabatic losses due to 2D conical expansion of the
shocked plasma sheets.  The main limitations of the present version of
the code are that Comptonization is treated self-consistently only
within the Thomson limits, and that $\gamma\gamma$-pair production is
not included.  Since there are observational indications that in FSRQ
the $\gamma$-ray spectra have a high energy break in the $4-10$GeV
band (Pohl et al.~\cite{pohl97}; Sikora et al.~\cite{sbmm02}), while
synchrotron spectra are usually steep in the UV band, the above
limitations affect only marginally our models of quasars.  However,
they can be significant for BL Lac objects, particularly for those
bright in the TeV band.

Assumptions and main features of our model, such as the equations
describing evolution of the energy distribution of relativistic
electrons, radiation processes, relativistic aberration and light
travel effects are presented in Sect.~2.  The numerical code is
described in Sect.~3. Approximate analytical formulas, expressing the
model input parameters as a function of observables, are derived in
Sect.~4.  In Sect.~5 we present results of application of our code to
model the outburst of 3C279, the best studied quasar in the
$\gamma$-ray band. The work is summarized in Sect.~6.
 
\section{The Model}
In our model a source of the non-thermal radiation is assumed to be a
thin shell, propagating down the conical jet with a constant speed,
and radiation is produced by relativistic electrons/positrons,
injected into a shell within a given distance range. This picture
approximates the internal shock scenario, where shocks are formed
following collisions of inhomogeneities moving with different radial
velocities (Sikora et al.~\cite{sbr94}; Spada et al.~\cite{spa01}).
Structure and dynamics of such internal shocks is in general very
complex and depends on a number of parameters, such as the relative
velocity of inhomogeneities, their densities, temperatures, geometry,
and total masses.  The shock structure can be double, single (forward
or reverse), or can initially it can be double, followed by an
evolution into a single shock (Daigne \& Mochkovitch~\cite{dm98};
Bicknell \& Wagner~\cite{bw02}).  Regardless, in the case of
inhomogeneities which are cold prior to their collision and which have
comparable masses and comparable rest frame densities, the constant
speed of the shocked plasma is a good approximation.  Small thickness
of the shocked plasma shell can be justified, since internal shocks
are at most mildly relativistic (Sikora et al.~\cite{sbbm01}).

\subsection{Electron Evolution}
Assuming that electron injection function and energy densities of
magnetic field and of external radiation fields are uniform across the
shell, one can follow evolution of electron energy distribution by
solving the kinetic equation for the total population of relativistic
electrons, despite the fact that each element of the conically
diverging shell has its own rest frame which is different from other
frames.  That equation can be written in the form (Moderski et
al.~\cite{msb00}):
\be
{\partial N_{\gamma} \over \partial t'} = - {\partial \over \partial
\gamma} \left(N_{\gamma} {{\rm d}\gamma \over {\rm d}t'}\right) + Q ,
\label{dNdtp}
\ee
where $Q \equiv (dN_{\gamma}/dt')_{\rm inj}$ is the electron injection
function, $\gamma$ is the random electron Lorentz factor,
$d\gamma/dt'$ is the rate of the electron energy losses, ${\rm d}t' =
{\rm d}t /\Gamma$ is the proper time, and $\Gamma$ is the Lorentz
factor of the bulk motion of a shell.  Within the thin shell
approximation, the time derivatives can be replaced with derivatives
over the radial distance $r$. Using a relation ${\rm d}r =
\beta_{\Gamma} c \Gamma {\rm d}t'$, where $\beta_{\Gamma} =
\sqrt{\Gamma^2-1}/\Gamma$, we obtain
\be
{\partial N_{\gamma} \over \partial r} = - {\partial \over \partial
\gamma} \left(N_{\gamma} {{\rm d}\gamma \over {\rm d}r}\right) + {Q
\over c \beta_{\Gamma} \Gamma},
\label{dNdr}
\ee
where
\be
{{\rm d}\gamma \over {\rm d}r} = {1 \over \beta_{\Gamma} c \Gamma}
\left({\rm d}\gamma \over {\rm d}t'\right)_{\rm rad}- {2 \over
3}{\gamma \over r} .
\label{dgdr}
\ee
The second term on the right-hand side of Eq.~(\ref{dgdr}) represents
the adiabatic losses due to two-dimensional (conical) expansion of the
shell.

\subsection{Radiative Cooling Rates}
There are three processes contributing to the radiative energy losses
of electrons:

\noindent
synchrotron radiation,
\be
\left({\rm d}\gamma \over {\rm d}t'\right)_{\rm S} = - {4\sigma_T \over
3 m_{\rm e} c } u_{\rm B}' (\beta \gamma)^2 ;
\label{dgdtS}
\ee
\noindent
Comptonization of synchrotron radiation,
\be
\left({\rm d} \gamma \over {\rm d} t'\right)_{\rm SSC} = - {4 \sigma_T
\over 3 m_{\rm e} c } u_{\rm S}' (\beta \gamma)^2 ;
\label{dgdtSSC}
\ee
and Comptonization of external radiation,
\be
\left({\rm d}\gamma \over {\rm d}t'\right)_{\rm ERC} = - {4 \sigma_T
\over 3 m_{\rm e} c } u'_{\rm ext} 
(\beta \gamma)^2   ,
\label{dgdtERC}
\ee
where $u'_{\rm B} = B'^2/8\pi$ is the magnetic energy density,
$u'_{\rm S}$ is the energy density of the synchrotron radiation field,
and $u'_{\rm ext}$ is the energy density of the external radiation
field, all as measured in the source comoving frame. $B'$ is the
strength of the magnetic field, which can be an arbitrary function of
the radius.

Energy density of the synchrotron radiation produced by a thin shell
is
\be
u_{\rm S}' \simeq {(t'-t_0')\delta L_{\rm S}' \over \delta V'} \simeq
{(t'-t_0')\delta L_{\rm S}' \over \lambda' r^2 \delta \Omega_{\rm j}} \simeq
{\delta L_{\rm S}' \over 2 c r^2 \delta \Omega_{\rm j}}
\ee
where $t'-t_0'$ is the time that elapsed from the beginning of the
shock formation, $\lambda' \simeq 2 c (t'-t_0')$ is the width of the
synchrotron radiation layer at that time, and
\be
\delta L'_{\rm S} = \int_{\nu_{\rm S,min}'}^{\nu_{\rm S,max}'} \, \delta
L_{{\rm S},\nu'}' {\rm d} \nu' \, ,
\label{us}
\ee
is the synchrotron luminosity produced in a cell of the shell,
enclosed within a solid angle $\delta \Omega_{\rm j} \ll \Omega_{\rm
j} \equiv \pi \psi_j^2$ (see Fig. 1).  The monochromatic synchrotron
luminosity, $\delta L_{{\rm S},\nu'}'$, is computed using
Eq.~(\ref{LS}) and integrated over the frequency range from $\nu_{\rm
S,min}' \simeq \nu'_{\rm abs}$, at which the optical thickness due to
the synchrotron self-absorption is equal to unity, to $\nu_{\rm
S,max}' = (2e/ 3 \pi m_{\rm e} c) \gamma_{\rm max}^2 B'$.

Energy density of the external radiation field is
\bea
\lefteqn { u_{\rm ext}' = {1 \over c} \int {I_{\rm ext}' {\rm
d}\Omega'} = {1 \over c} \int {{I_{\rm ext} \over {\cal D}_{\rm in}^2}
{\rm d}\Omega} \simeq \null} \nonumber \\ && \null \simeq {\Gamma^2
\over c} \int {I_{\rm ext} (1-\beta_{\Gamma}\cos{\theta_{\rm in}})^2
{\rm d} \Omega} \,
\label{uext}
\eea
where $\theta_{\rm in}$ is the angle between the trajectory of the
incoming photon and the direction of the cell motion. For narrow jets
this angle can be approximated by the angle between the trajectory of
the incoming photon and the jet axis.

\subsection{Intrinsic Luminosities}

\subsubsection{Synchrotron Radiation}
The rate of the synchrotron radiation production per cell for a given
electron distribution is (Chiaberge \& Ghisellini~\cite{cg99}):
\be
\delta L_{{\rm S},\nu'}'(r) = \int_{\gamma_{\rm min}}^{\gamma_{\rm max}}
\delta N_{\gamma}(r) F_{\rm S}(\nu',\gamma) {\rm d} \gamma \, ,
\label{LS}
\ee
where $\delta N_{\gamma} = N_{\gamma} (\delta \Omega_{\rm
  j}/\Omega_{\rm j})$, and
\bea
\lefteqn {F_{\rm S}(\nu',\gamma) = {3 \sqrt{3} \over \pi} {\sigma_T c
u_{\rm B}' \over \nu_{\rm B}'} \chi^2 \times \null} \nonumber \\ 
&& \null \times \left \{ K_{4/3}(\chi) K_{1/3}(\chi) - {3 \over 5} \chi
\left [ K_{4/3}^2(\chi) - K_{1/3}^2(\chi) \right ] \right \}
\eea
is the single electron synchrotron emissivity averaged over an
isotropic distribution of pitch angles (Crusius \&
Schlickeiser~\cite{cs86}), where $\chi=\nu'/(3 \gamma^2 \nu'_{\rm B})$
and $\nu_{\rm B}' = e B'/(2 \pi m_{\rm e} c)$.  The power $F_{\rm
S}(\nu',\gamma)$ is taken with the low energy cutoff at $\nu_{\rm
abs}'$.

Note that because the time scale of the electron/po\-sitron gyration in
the local tangled magnetic fields is much shorter than the time scale
of radiative energy losses (Dermer \& Schlickeiser~\cite{ds93}), we
can assume that the electron momentum distribution is isotropic in the
cell co-moving frame.  It does not imply that the synchrotron
radiation field is isotropic.  In general, one must transform the
synchrotron radiation from neighbor cells into cell frame taking into
account the change in frequency and direction of the incoming photons.
However, for a thin shell and small opening angles of the jet the
assumption about isotropy of the synchrotron radiation field is a
reasonable approximation.

\subsubsection{Synchrotron-Self-Compton}
Using $\delta$-function approximation, one can find (Chiang \&
Dermer~\cite{chd99})
\bea
\lefteqn{\delta L_{{\rm SSC},\nu'}' (r) \simeq {\sqrt 3 h c
\sigma_{\rm T} \over 4} {\nu'}^{1/2} \times \null} \nonumber \\ &&
\times \int_{\nu_1'}^{\nu_2'} \delta N_{\gamma}(r) n_{{\rm S},\nu_{\rm
S}'}' \nu_{\rm S}'^{-1/2} \, {\rm d} \nu_{\rm S}' \, ,
\label{LSSC}
\eea
where
\be
\nu_1' = {\rm Max} \left [ \nu_{\rm abs}'; {3 \nu' \over 4
\gamma_{\rm max}^2} \right ] \, ,
\label{nu1}
\ee
and
\be
\nu_2' = {\rm Min} \left [ \nu_{\rm S, max}'; {3 \nu' \over 4
\gamma_{\rm min}^2} ; {3 (m_{\rm e} c^2)^2 \over 4 h^2 \nu'} \right ] \, ,
\label{nu2}
\ee
where $n_{{\rm S},\nu_{\rm S}'}'$ is the number density of the
synchrotron photons per frequency.  Noting that for isotropic
synchrotron radiation field $n_{{\rm S},\nu'}' = u_{{\rm S},\nu'}' /h
\nu'$ and
\be
u_{{\rm S}, \nu'}' \simeq
{\delta  L_{{\rm S},\nu'}' \over 2  r^2 \delta \Omega_{\rm j} c} \, 
\label{usm}
\ee
(see Eq.~(\ref{us})), we finally obtain
\bea
\lefteqn{\delta L_{{\rm SSC}, \nu'}' = {\sqrt 3 \sigma_{\rm T} \over 8 r^2
\delta \Omega_{\rm j}} {\nu'}^{1/2} \times \null} \nonumber \\ && \null
\times \int_{\nu_1'}^{\nu_2'} \delta N_{\gamma} \left [\gamma= \sqrt
{3 \nu' \over 4 \nu_{\rm S}'} \, \right ] \delta L_{{\rm S},\nu'}'
\nu_{\rm S}'^{-3/2} \, {\rm d} \nu_{\rm S}' \, .
\label{LSSCbis}
\eea
This is analogous to the Eq.~(27) of Chiang \&
Der\-mer~(\cite{chd99}).

\subsubsection{External-Radiation-Compton}
In the co-moving frame, the external diffuse radiation field is
strongly anisotropic.  Due to relativistic aberration, the external
radiation appears to the jet as mostly incident from the forward
direction.  Within the jet, a photon scattered by a relativistic
electron follows the direction of motion of that electron.  Because of
this, the observer will detect preferentially photons that were
emitted by the electrons which during the scattering process had their
momentum vector pointing at the observer.  In such an approximation,
the energy of the photon scattered into the observer direction is
given by (Reynolds~\cite{rey82}):
\be
\nu' \simeq \gamma^2 \nu_{\rm ext}' (1 + \cos \theta') \, ,
\label{nup}
\ee
where $\nu_{\rm ext}' \simeq \Gamma \nu_{\rm ext}$, $h \nu_{\rm ext}$
is the characteristic energy of photons in the external radiation
field (assumed to have a narrow spectrum), and $\theta'$ is the angle
between the bulk motion direction of a given shell segment and
direction to the observer, as measured in the co-moving frame of the
bulk motion.  For $\gamma \gg 1$ the rate of energy losses of an
electron with a velocity vector instantaneously oriented into
direction of the observer is:
\be
\left \vert {\rm d} \gamma \over {\rm d} t' \right \vert_{\rm ERC} [\theta']
\simeq  { \sigma_{\rm T} \over  m_{\rm e} c} u_{\rm ext}' (1 + \cos \theta')^2
\gamma^2 \, .
\label{dgdtth}
\ee
Note that the angle-averaged value of $({\rm d} \gamma / {\rm d}
t')_{\rm ERC} [\theta']$ gives Eq.~(\ref{dgdtERC}) with $\beta \simeq
1$.  Since the ``blazar phenomenon'' implies observing the jet at very
small angles, $\theta \ll \pi/2$, the co-moving observing angle
$\theta'$ is much smaller than $\pi - 1/\Gamma$ and then
(Dermer~\cite{der95})
\be
\Gamma (1 + \cos \theta') \simeq \Gamma (1 + \beta_{\Gamma} \cos
\theta') = {\cal D} \, , \label{aber}
\ee
With that approximation the Eqs~(\ref{nup}) and (\ref{dgdtth}) can be
rewritten in the form
\be
\nu' \simeq {\cal D} \gamma^2 \nu_{\rm ext} \, ,
\label{nubis}
\ee
and
\be
\left \vert{\rm d} \gamma \over {\rm d} t'\right \vert_{\rm ERC} [\theta']
\simeq {\sigma_{\rm T} \over  m_{\rm e} c} u_{\rm ext}' \gamma^2 
\left({{\cal D}\over \Gamma }\right)^2  \, .
\label{dgdtthbis}
\ee

\noindent
The power per unit solid angle and frequency, emitted by a cell via
the ERC process into the observer direction is, as measured in the
cell co-moving frame,
\be
{\partial \delta L_{{\rm ERC},\nu'}'[\theta']
\over \partial \Omega_{{\vec n}_{\rm obs}}'} \simeq {\delta N_{\gamma}
\over 4 \pi} \, \left \vert {\rm d} \gamma \over {\rm d} t' \right
\vert_{\rm ERC} [\theta'] m_{\rm e} c^2 {{\rm d} \gamma \over {\rm d} \nu'}
\, ,
\label{LERCp}
\ee
and after combining with Eqs~(\ref{nubis}), (\ref{dgdtthbis}), and
(\ref{uext}), we obtain
\bea
\lefteqn{ {\partial \delta L_{{\rm ERC},\nu'}'[\theta'] \over
\partial \Omega_{{\vec n}_{\rm obs}}'} \simeq {\sigma_{\rm T}
(\gamma \delta N_{\gamma}) {\cal D} \over 8 \pi \nu_{\rm ext}} \times
\null} \nonumber \\ && \null \times \int{
(1-\beta_{\Gamma}\cos{\theta_{\rm in}})^2 \, I_{\rm ext} \, {\rm d}
\Omega} \, .
\label{LERCpbis}
\eea

\subsection{Observed Spectra and Light Curves}
The monochromatic radiation flux observed at a given instant is
\be
F_{\nu_{\rm obs}}(t_{\rm obs}) = {(1+z) \over  d_{\rm L}^2} \,
\sum {\partial \delta L_{\nu} [t, \theta] \over 
\partial \Omega_{{\vec n}_{\rm obs}}} \, 
\label{flux}
\ee
where
\be
{\partial \delta L_{\nu} [t, \theta] \over 
\partial \Omega_{{\vec n}_{\rm obs}}} = {\cal D}^3
{\partial \delta L_{\nu'}' [r, \theta'] \over 
\partial \Omega_{{\vec n}_{\rm obs}}'} \, ,
\label{Lnu}
\ee
\be 
\nu' = \nu /{\cal D} = \nu_{\rm obs}(1+z)/{\cal D} \, , 
\ee
\be
r = {c \beta_{\Gamma} \over 1 - \beta_{\Gamma} \cos \theta} (t-t_0)
+ r_0 \, ,
\label{rt}
\ee
\be 
\cos \theta = {\cos \theta' + \beta_{\Gamma} \over 1 +
\beta_{\Gamma}\cos \theta'} \, ,
\ee
\be
t=t_{\rm obs}/(1+z) \, ,
\ee
\be {\cal D}= \Gamma (1 + \beta_{\Gamma} \cos{\theta'}) \equiv
{1 \over \Gamma (1 - \beta_{\Gamma} \cos{\theta})} \, \ee
and $d_L$ is the luminosity distance. For the $\Omega_{\lambda} +
\Omega_m = 1$ cosmology,
\be
d_{\rm L} = {c \over H}(1+z) \int_0^z {{dx \over [\Omega_{\lambda} +
\Omega_{\rm m}(1+x)^3]^{1/2}}} \, .
\ee
where $H$ is the Hubble constant.

It should be noted here that because the synchrotron and SSC emission
is isotropic in the cell co-moving frame, the co-moving power per
solid angle produced by those processes is
\be
{\partial \delta L_{{\rm S,SSC},\nu'}'[\theta'] \over \partial
\Omega_{{\vec n}_{\rm obs}}'} = {\delta L_{\rm S,SSC,\nu'}' \over 4
\pi} \, . 
\ee
where $\delta L_{\rm S,\nu'}$ and $\delta L_{\rm SSC,\nu'}$ are given
by Eqs~(\ref{LS}) and (\ref{LSSC}), respectively.

\section{Numerical Implementation}
Our numerical method is similar to that used by Chia\-berge \&
Ghisellini~(\cite{cg99}).  Electron evolution equation (\ref{dNdr}) is
solved with implicit difference scheme adopted from Chang \&
Cooper~(\cite{cc70}).  First, the uniformly spaced logarithmic energy
grid is established:
\be
\gamma_k = \gamma_0 \left ( {\gamma_N \over \gamma_0} \right )^{k - 1
\over n - 1} ~~~~~~ k = 0,...,n
\label{grid}
\ee
with energy intervals $\Delta \gamma_k = \gamma_{k+1/2} -
\gamma_{k-1/2}$.  Using a definition $N^i_k = N_\gamma(\gamma_k,i
\Delta r)$ we can convert Eq.~(\ref{dNdr}) into a set of linear
equations:
\be
A_k N^{i+1}_{k-1} + B_k N^{i+1}_k + C_k N^{i+1}_{k+1} = U^i_k ,
\label{set}
\ee
where the coefficients are
\be
\begin{array}{l}
A_k = 0 \, ,\\
B_k = 1 + \left ( {{\rm d} \gamma \over {\rm d} r} \right )_{k-1/2}
{\Delta r \over \Delta \gamma_k} \, ,\\
C_k = - \left ( {{\rm d} \gamma \over {\rm d} r} \right )_{k+1/2}
{\Delta r \over \Delta \gamma_k} \, ,\\
\end{array}
\ee
and the source term is
\be
U^i_k = N^i_k + {\Delta r \over c \beta_\Gamma \Gamma} Q^i_k \, .
\ee

\noindent
The system of equations~(\ref{set}) yields a special case of
tridiagonal matrix with the upper diagonal equal to zero.  It is
solved numerically by matrix decomposition, forward- and
back-substitution.  In practice we use {\tt tridag} subroutine from
Numerical Recipes (Press et al.~\cite{ptvf92}). This method of solving
Eq.~(\ref{dNdr}) is unstable if the sharp cutoffs are present in
the electron energy distribution function.  In our case, because the
Eq.~(\ref{dgdr}) does not contain the acceleration terms, we
avoid this instability by setting the upper boundary of the
calculation grid to the maximum energy of the injected
electrons. Another way to avoid sharp cutoffs is to introduce very
steep power-law tails to mimic these cutoffs (B\"ottcher \&
Chiang~\cite{bch02}).

\noindent
There exists an alternative way of solving Eqs~(\ref{set}).  If
one assures that the energy grid is sufficiently wide to have
$N(\gamma_n,i \Delta r) = 0$ during the entire evolution, then the
Eq.~(\ref{set}) can be solved recursively from the highest to the
lowest energies using the relation
\be
N^{i+1}_k = {1 \over B_k} \left ( U^i_k - C_k N^{i+1}_{k+1} \right
) ~~~~~ k = n-1, ... ,1
\ee
For universality, however, we use the full {\tt tridag} method.  It
allows the consideration of electron distributions with high energy
tails without performing the calculations for unimportant grid points.

\noindent
In order to calculate the electron distribution it is necessary to
estimate the electron cooling function~(\ref{dgdr}).  This requires
the knowledge of the energy density of the synchrotron radiation field
which depends on the electron distribution itself.  In our code this
calculation is done iteratively.  Initially $u'_{\rm S} = 0$ is
assumed and the electron distribution is calculated using this value.
Then this distribution is used to calculate synchrotron radiation
density together with the frequency of synchrotron self
absorption. The new value of $u'_{\rm S}$ is used to recalculate the
electron distribution.  This process is repeated until convergence.

\noindent
The electron distribution is then used to calculate the synchrotron
luminosity (\ref{LS}), SSC luminosity~(\ref{LSSC}), and ERC
luminosity~(\ref{LERCp}).  For all integrations we use the Romberg's
method (Press et al.~\cite{ptvf92}).  The advantage of this method is
that it adjusts itself, and thus only a minimal number of calculations
is performed to achieve desired accuracy.

\noindent
Special care must be taken when calculating light curves from
Eq.~(\ref{flux}).  The radiation reaching the observer at a given
time is a superposition of radiation emitted at different radii, and
this in turn depends on $\theta$, which is the angle between the
direction of motion of a given cell and the direction to the observer
(see Eq.~(\ref{rt})).

\noindent
One must also remember that the choice of the cell size puts a
constraint on the minimum timescale that can be probed with the code.
For a given cell size, $\Delta \theta$, the minimum timescale for
which spatial homogeneity within the cell is maintained equals
\be
\Delta t \simeq r \sin \theta_{\rm j} \Delta \theta
\ee

\begin{figure}
\resizebox{\hsize}{!}{\includegraphics{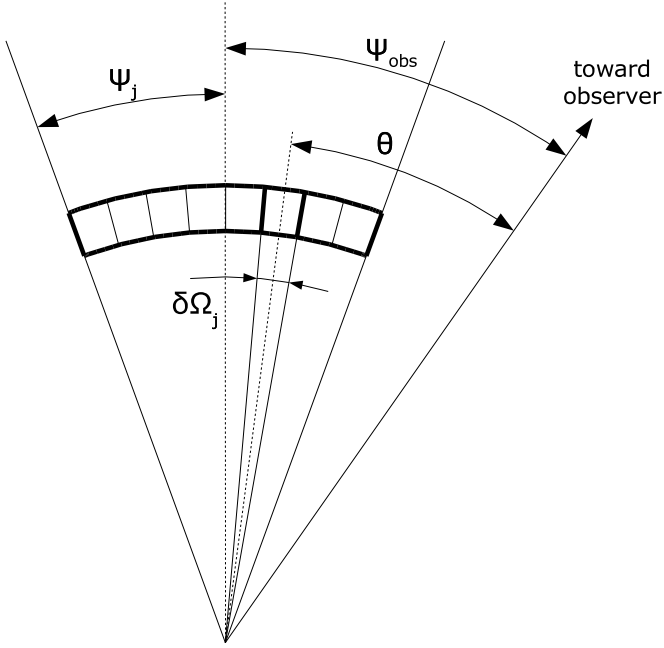}}
\caption{ 
  A schematic picture illustrating the integration
  procedure. A thin shell of electrons enclosed within the
  half-opening angle of the jet, $\psi_{\rm j}$ is divided into cells,
  each $\delta \Omega_{\rm j}$ thick. Each such cell contains $\delta
  N_\gamma$ electrons. The observer is located at the angle $\psi_{\rm
  obs}$ from the jet axis. Integration is performed by summation over
  $\theta$ contributions from cells located at different radii.}
\label{figjet}
\end{figure}

\section{Model Input Parameters}

The input parameters of the model are: 

\noindent
-- the distance of the shock formation from the central source, $r_0$;

\noindent
-- the distance range of shock operation, $\Delta r_{\rm coll}$;

\noindent
-- the bulk Lorentz factor of the shock, $\Gamma$;

\noindent
-- the energy density of an external diffuse radiation field, 
$u_{\rm ext}$ and its characteristic frequency $\nu_{\rm ext}$;

\noindent
-- the magnetic field in a shocked plasma, $B'$;

\noindent
-- the electron injection function, $Q$; 

\noindent
-- the angle between the line of sight and the jet axis, 
$\psi_{\rm obs}$;

\noindent
-- the half-opening angle of a jet, $\psi_{\rm j}$.

Initial values of these parameters can be set up by using approximate
analytical formulas which relate them to the observable quantities.
Depending on the available observables, a different set of equations
must be used.  Most useful observables used to model the short term
($1-10$ days), $\gamma$-ray dominated outbursts in FSRQ are: the time
scale of the outburst/flare, $t_{\rm fl}$ (more specifically, its
growing part); radiation fluxes at the low-energy and high-energy
spectral peaks, denoted respectively as $F_{\rm LE} \equiv \nu_{\rm
LE} F_{\nu_{\rm LE}}$ and $F_{\rm HE} \equiv \nu_{\rm HE} F_{\nu_{\rm
HE}}$; location of the spectral break between the X-ray and
$\gamma$-ray spectral portions, $\nu_{\rm c}$; radiation flux,
$\nu_{\star}F_{\nu_{\star}}$, at some $\nu_{\star} > \nu_{\rm c}$; the
spectral index of the HE component at $\nu > \nu_{\rm c}$ (as
measured, e.g., by EGRET in its bandpass); and the total radiation
flux of broad emission lines, $F_{\rm BEL}$ (Celotti et
al.~\cite{cpg97}).  For the sake of simplicity of analytical formulas
which will be used to derive the initial model parameters as a
function of observables, both the light travel effects (due to finite
size of the source) and the Doppler dispersion effects (due to the
``conical'' expansion of a shell) are ignored. In addition, we assume
that: during the outbursts the production of $\gamma$-rays is
dominated by the ERC process; $B' \propto 1/r$; and electrons are
accelerated in the shock at a constant rate and with a single
power-law energy distribution,
\be
Q=K \gamma^{-p} \, .
\label{Q}
\ee
for $\gamma_{\rm min}=1$ and  
\be 
\gamma_{\rm max}= {1 \over {\cal D}} \sqrt{\nu_{\rm max}(1+z) \over
  \nu_{\rm BEL}}
\label{gm}
\ee
(see Appendix~A  for justification of such an injection function).

Due to radiative and adiabatic energy losses of electrons, their
energy distribution evolves with time. The strong energy dependence of
the radiative energy losses of electrons causes steepening of the
electron energy distribution. It takes place at $\gamma > \gamma_{\rm c}$,
where $\gamma_{\rm c}$ is the energy at which the time scale of electron
energy losses,
\be
t_{\rm cool}' \simeq {\gamma \over \vert d \gamma/dt'\vert}  \simeq 
{\gamma \over \vert d \gamma/dt'\vert_{\rm rad} + \vert d
  \gamma/dt'\vert_{\rm ad}}
\, ,
\label{tc}
\ee
is equal to the lifetime of a shock
\be
t_{\rm sh}' \simeq {t_{\rm fl} {\cal D} \over 1+z} \, .     
\ee 
During $\gamma$-ray dominated outbursts the radiative energy losses
are dominated by the ERC process, and therefore $t_{cool}'=t_{sh}'$
gives
\be
\gamma_{\rm c} = {m_{\rm e} c \, (1+z) f(k) \over 2 \sigma_{\rm T}
t_{\rm fl} u_{\rm ext}' {\cal D}} \, ,
\label{gc}
\ee
where
$f(k) =(1+3k)/ (2(1+k))$ and $k\equiv r_0/\Delta r_{\rm coll}$.

At $\gamma =\gamma_{\rm c}$ the power law energy distribution of electrons,
$N_{\gamma} \propto \gamma^{-s}$, changes the slope from $s=p$ at
$\gamma < \gamma_{\rm c}$ (slow cooling regime) to $s=p+1$ at $\gamma >
\gamma_{\rm c}$ (fast cooling regime). Such a change of slope results from
the fact that in the fast cooling regime ($t_{\rm rad} < t_{\rm sh}$) the
number of electrons is saturated by the radiative losses, i.e.
$N_{\gamma} \sim \int_{\gamma} Q \, d\gamma / \vert d\gamma/dt'\vert
\sim Q t_{\rm rad}'$.  (Note that $\gamma_{\rm c}$ is changing during the shock
progression and that its particular value given by Eq.~(\ref{gc}) is
for the instant when the collision is complete.)
 
The break at $\gamma_{\rm c}$ is reflected in the ERC spectral
component at
\be
\nu_{\rm c,obs} = {\gamma_c^2 {\cal D}^2 \nu_{\rm ext} \over 1+z}  \, .
\label{nuc}
\ee 
around which the electromagnetic spectrum changes the slope by $\Delta
\alpha \simeq 0.5$. We identify this break as the one observed more or
less directly in FSRQ between 1 and 30 MeV. Such identification
provides information about energy density of external radiation field
as measured in the shock co-moving frame.  Using Eqs~(\ref{gc}) and
(\ref{nuc}), we obtain
\be
u_{\rm ext}' (r_{\rm f})= { m_{\rm e} c \over 2 \sigma_{\rm T}}
{(1+z)^{1/2} f(k) \over t_{\rm fl} (\nu_{\rm c}/\nu_{\rm ext})^{1/2}} \,
\label{uex1} ,
\ee
where $r_{\rm f} \simeq r_0 + \Delta r_{\rm coll}$.  Since
observations provide directly the ratio $F_{\rm HE}/F_{\rm LE}$ and
since
\be
{F_{\rm HE} \over F_{\rm LE}} =
{F_{\rm ERC} \over F_{\rm S}} \simeq 
{ {\vert {\rm d}\gamma /{\rm d} t'\vert }_{\rm ERC}[\theta']
\over {\vert {\rm d}\gamma /{\rm d} t'\vert }_{\rm S}} \simeq 
{u_{\rm ext}' \over u_{\rm B}'}
\left ({{\cal D} \over \Gamma} \right)^2  \, ,
\label{ff}
\ee
we can estimate the intensity of the magnetic field,
\be
B'(r_{\rm f}) \simeq {{\cal D} \over \Gamma}
\, \sqrt{8\pi u_{\rm ext}' {F_{\rm LE} \over F_{\rm HE}}}   \, ,
\label{B2}
\ee
where $u_{\rm ext}'$ is given by Eq.~(\ref{uex1}).

The short term flares (lasting typically 1 -- 10 days) are most likely
produced at distances $0.1 - 1.0$ pc. At such distances, the external
photon energy density $u_{\rm ext}'$ is dominated by broad emission
lines (see, e.g., Sikora et al.~\cite{sbmm02}).  With this, and
assuming spherical geometry of the BEL region, we have
\be
u_{\rm ext}'(r) \simeq 
{4 \over 3} \Gamma^2 
{(\partial L_{\rm BEL}/\partial \ln r)(r) \over 4\pi r^2 c }\, .
\label{uex2}
\ee
Provided that $(\partial L_{\rm BEL}/\partial \ln r)(r)$ is known, this
equation, combined with Eq.~(\ref{uex1}) and with the formula for the
distance range of the shock operation,
\be 
\Delta r_{\rm coll} \simeq {c t_{\rm fl} \over 1 - \beta_{\Gamma}
\cos\psi_{\rm obs}} \, {1 \over 1+z} \equiv {c t_{\rm fl} \Gamma^2 \over
1+z}{{\cal D}\over \Gamma} \, ,
\label{Del}
\ee
can be used to estimate $r_{\rm f}$ and $\Gamma$.  For a broad emission line
luminosity distribution approximated by (see Appendix~\ref{app2})
\be
{\partial L_{\rm BEL} \over \partial \ln r} (r) = {L_{\rm BEL}
\over 4} \times \cases{ (r/r_{\rm BEL})^{0.5}& for $r < r_{\rm BEL}$
\cr (r/r_{\rm BEL})^{-0.5} & for $r > r_{\rm BEL}$ \cr} \,
\ee
where
\be
r_{\rm BEL} \simeq 10^{18} (L_{\rm UV}/10^{46}{\rm ergs
\, s}^{-1})^{0.7} \, {\rm cm} \, ,
\label{rbel}
\ee
is the distance at which the radial distribution of the BEL luminosity
has a maximum (Kaspi et al.~\cite{kas00}), we obtain
\be
r_{\rm f} \simeq \cases{ c_{\rm r}^2 (\nu_{\rm c}/\nu_{\rm BEL})
L_{\rm BEL}^2 r_{\rm BEL}^{-1} & if $r_f < r_{\rm BEL}$ \cr c_{\rm
r}^{2/3} (\nu_{\rm c}/\nu_{\rm BEL})^{1/3} L_{\rm BEL}^{2/3} r_{\rm BEL}^{1/3}
& if $r_{\rm f} > r_{\rm BEL}$ \cr} \, ,
\label{rf}  
\ee
where $\nu_{\rm BEL}$ is a characteristic frequency of the broad emission
lines radiation, and  
\be
\Gamma = \sqrt{{(1+z) r_{\rm f} \over (1+k) c t_{\rm fl} ({\cal
D}/\Gamma)}} \, .
\label{Gamma}
\ee
where $c_{\rm r} = \sigma_{\rm T} (1+z)^{1/2}/[6\pi m_{\rm e} c^3
(1+k)f(k)({\cal D}/\Gamma)]$, $L_{\rm BEL} = 4\pi d_{\rm L}^2 F_{\rm
BEL}$, and $L_{\rm UV} = 4\pi d_{\rm L}^2 F_{\rm UV}$ is the
luminosity of the accretion disc.  The disc radiation is sometimes
observed directly (during low states of FSRQ); otherwise it can be
estimated from $F_{\rm UV} \simeq F_{\rm BEL}/\xi_{\rm BEL}$, assuming
the canonical value of the BEL covering factor, $\xi_{\rm BEL} =0.1$.

In order to determine the parameters of the electron injection
function, we use $\gamma$-ray data from the spectral band at $\nu \gg
\nu_{\rm c}$.  There, deeply in the fast cooling regime,
\be
p=2 \alpha_{\gamma} \,
\label{indp}
\ee
and
\be
N_{\gamma} \simeq \left \vert{1 \over \dot \gamma}\right
\vert_{\rm tot} \int_{\gamma} Q \, d\gamma \, .
\ee
Noting that for $\gamma$-ray dominated outbursts, $\vert \dot
\gamma\vert_{\rm tot} \simeq \vert \dot \gamma\vert_{\rm ERC}$, by
using approximate formula for ERC radiation production,
\be
4 \pi d_{\rm L}^2 \nu_{\star} F_{\nu_{\star}} \simeq 0.5 [\gamma
N_{\gamma}] \vert \dot \gamma \vert_{\rm tot} m_{\rm e} c^2 \Gamma^4
({\cal D}/\Gamma)^6 \, ,
\ee
one can find that the normalization factor of the electron injection
function is
\be
K \simeq {8 \pi d_{\rm L}^2 [\nu_{\star}F_{\nu_{\star}}] \over m_{\rm e} c^2
\Gamma^4 ({\cal D}/\Gamma)^6} g(\alpha_{\gamma}) \, ,
\label{injK}
\ee
where
$g(\alpha_{\gamma}) =
(2\alpha_{\gamma}-1)\gamma_{\star}^{2(2\alpha_{\gamma}-2)}$
and
\be
\gamma_{\star} = {1 \over {\cal D}} \sqrt{
  \nu_{\star}(1+z) \over \nu_{\rm ext}}.
\ee
Note that for $\alpha_{\gamma} = 1$, $g(\alpha_{\gamma})= 1$.

To complete the set of the model input parameters we still need to
specify the values of ${\cal D}/\Gamma$, $k$ and $\psi_{\rm j}$. Since the
ERC to synchrotron luminosity ratio is $\propto ({\cal D}/\Gamma)^2$
(see Eq.~(\ref{ff})), the view angles $\psi_{\rm obs} \le 1/\Gamma$ are
expected for strongly $\gamma$-ray dominated FSRQ.  In such conditions
${\cal D}/\Gamma=1$ ranges from $2$ for the observer located on the
jet axis to $1$ for the observer located at $\psi_{\rm obs} = 1/\Gamma$.

The value of $k$ is also expected to be of the order of unity, as
suggested by roughly symmetrical profiles of flares (see, e.g., Sikora
et al.~\cite{sbbm01}). Regarding the jet opening angle, we know they
are very small (1 - 3 degrees) on kilo-parsec scales, but can be much
larger on parsec/sub-parsec scales (see, e.g. Lobanov~\cite{lob98}).
The specific value of $\psi_{\rm j}$ can be determined from the value
of flux in the soft/mid X-ray bands, provided that the latter is
dominated by the SSC process. Since we do not follow the SSC process
in our analytical approach, and because other model parameters are not
very sensitive to $\psi_{\rm j}$, the numerical simulations can be
started with any $0 <\psi_{\rm j} <1/\Gamma$ and then corrected in
subsequent iterations.

In our method to calculate the model input parameters, we do not take
advantage of the ratio of frequencies, $\nu_{\rm HE}$ to $\nu_{\rm
LE}$. This ratio often has been used to calculate the bulk Lorentz
factor. If it were to be incorporated into our method to calculate
$\Gamma$, this would give an equation for $k$. However, $\nu F_{\nu}$
spectra around their maxima are quite flat and therefore, both
$\nu_{\rm HE}$ and $\nu_{\rm LE}$ are subject to very large
uncertainties.  Furthermore, in FSRQ the synchrotron peak is located
in the far IR and rarely is observed directly.

An additional difficulty is the fact that the amplitudes of the
$\gamma$-ray flares are usually much larger than the amplitudes of the
synchrotron flares. This suggests that synchrotron component is
strongly diluted by the quasi-steady radiation produced at larger
distances in a jet.  We can deal with this case in our method by using
in the Eq.~(\ref{B2}) $\Delta F_{\rm syn}$ instead of $F_{\rm syn}$,
where $\Delta F_{\rm syn}$ is the amount by which synchrotron flux
increases during the high amplitude $\gamma$-ray flare.  For the
instances where the dilution is so large that $\Delta F_{\rm syn}$
cannot be determined observationally, as an alternative approach, one
can use the observed correlation of the soft/mid X-ray flux with the
$\gamma$-ray flux.  When the soft/mid X-ray flux is interpreted as
produced by the SSC process, one can estimate the magnetic field
intensity via an iterative process, which relies on the match of the
model SSC radiation spectrum against that measured in the X-ray band.

\section{Application: blazar 3C 279}
3C 279 is one of the most extensively observed blazars. Several times
it was a target of multi-wavelength campaigns. The most fruitful
campaign took place in the beginning of 1996 when the blazar underwent
an enormous $\gamma$-ray flare. This event was monitored nearly
simultaneously at many frequencies (Wehrle et~al.~\cite{weh98}) and is
analyzed below using our code.

\subsection{Input parameters}
3C279 is located at a redshift $z=0.538$, which for the currently
favored cosmology ($\Omega_\lambda \simeq 0.7$, $\Omega_{\rm m} \simeq
0.3$ and $h \simeq 0.66$) gives the luminosity distance $d_{\rm L}
\simeq 10^{28}$cm. The data collected during the February 1996 flare
(Wehrle et~al.~\cite{weh98}; Hartman et al.~\cite{har01}) give the
$h\nu_{\rm c} \simeq 20$MeV and $\nu_{\star}F_{\nu_{\star}} \simeq
2.75 \times 10^{-10}$erg~s$^{-1}$~cm$^{-2}$, where $\nu_{\star} \simeq
10^{23}$Hz.  The slope of the $\gamma$-ray spectrum during the flare
in the EGRET band was $\alpha_{\gamma} \simeq 0.97$ which from
Eq.~(\ref{indp}) gives $p \simeq 1.9$.  Although a reliable EGRET
spectrum is available for this object, the high energy break is not
directly observed, so we set $h\nu_{\rm max} = 20$GeV. Estimate of
$F_{\rm BEL}$ is very uncertain. Analysis based on relative line
intensities by Celotti et al.~(\cite{cpg97}) gives $L_{\rm BEL} = 3.4
\times 10^{44}$erg s$^{-1}$.  However, during outbursts this
luminosity can be even larger (Koratkar et al.~\cite{kor98}).  For our
analysis we set $L_{\rm BEL} = 6.8 \times 10^{44}$erg s$^{-1}$. The
average photon energy of external radiation field is taken to be $h
\nu_{\rm ext} = h \nu_{\rm BEL}=10$eV.  Luminosity of the accretion
disc, obtained using the direct observations of the UV bump during the
low state of 3C 279 (Pian et al.~\cite{pian99}), is $L_{\rm UV} \simeq
2.7 \times 10^{45}$ erg s$^{-1}$.  Time scale of the flare is $ t_{\rm
fl} \simeq 1$ day (Lawson et al.~\cite{lhm99}).

Analysis of the lightcurves of the February 1996 flare at different
energies (Wehrle et~al.~\cite{weh98}) shows that despite the very
large rise of flux in the $\gamma$-ray band, the flare was almost
undetectable at low energies, especially from radio to optical. This
suggests that synchrotron component is significantly diluted, and
synchrotron radiation produced during the flare is hidden by a more
steady component. For this reason we initially set $F_{\rm syn}/F_{\rm
ERC} = 10^{-2}$.  The above set of observables yields (for the assumed
$k=1$ and ${\cal D}/\Gamma = 2$) the following input parameters:
$u_{\rm ext}'(r_{\rm f}) \simeq 0.21$erg cm$^{-3}$ ($\leftarrow$
Eq.~(\ref{uex1})); $B'(r_{\rm f}) \simeq 0.46$ Gauss ($\leftarrow$
Eq.~(\ref{B2})); $r_{\rm BEL} \simeq 4.0 \times 10^{17}$ cm
($\leftarrow$ Eq.~(\ref{rbel})); $r_{\rm f} \simeq 4.2 \times 10^{17}$
cm ($\leftarrow$ Eq.~(\ref{rf})); $\Gamma \simeq 7.9$ ($\leftarrow$
Eq.~(\ref{Gamma})); $K \simeq 3.4 \times 10^{48}$ s$^{-1}$
($\leftarrow$ Eq.~(\ref{injK})); $\gamma_{\rm max} \simeq 3.5 \times
10^3$ ($\leftarrow$ Eq.(~\ref{gm})).  We have also set $\psi_{\rm j} =
0.12 \simeq 1/\Gamma$.

\subsection{Average spectrum}
\begin{figure}
\resizebox{\hsize}{!}{\includegraphics{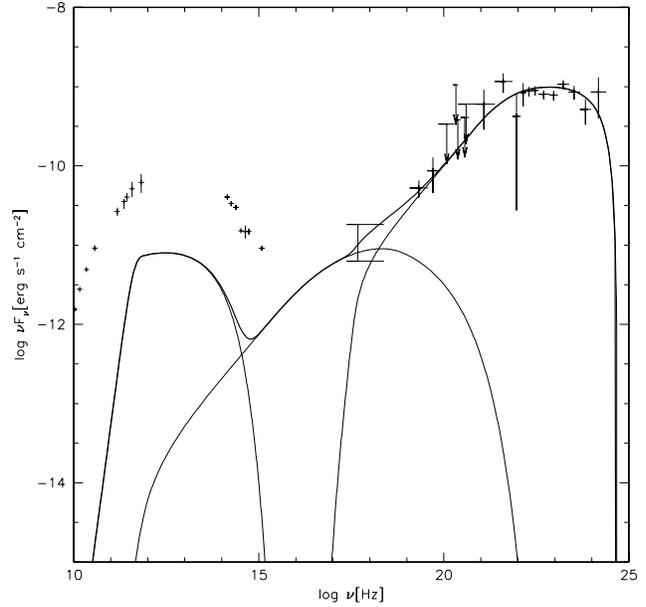}}
\caption{ 
  The average spectrum of the blazar 3C~279 during the
  February 1996 flare. Data points are from Hartman
  et~al.~(\cite{har01}). Thick, solid line shows the averaged spectrum
  of our model (see text for parameters). Thin lines represent various
  components of the spectrum.}
\label{spec}
\end{figure}
We first apply our model to the average spectrum of the blazar 3C~279
during the February 1996 flare. The result is presented in
Fig.~\ref{spec}. The fine-tuning of the input parameters requires
substantial change only of of the normalization of electron injection
function, $K$. We have achieved the best results setting $K = 9.5
\times 10^{49}$. This difference can easily be understood by examining
Eq.~(\ref{injK}). This quantity is very sensitive to ${\cal
D}/\Gamma$. The calculation of the input parameters assumes that the
source is point-like and thus that the whole radiation is beamed
toward the observer. Due to the conical structure of the jet the
Doppler factor decreases to the edge of the jet, and thus more
electrons are required to produce the observed luminosity. The minor
correction was applied also for the magnetic field. The best result
yields $B'(r_{\rm f}) = 0.3$ Gauss.  For the calculation we also set
$\psi_{\rm obs} = 0.05$, but the averaged spectrum is not very
sensitive to this value.  The averaging was performed between $r =
r_0$ and $r = 3r_0$.

Large discrepancy between the data and our model in the low energy
component arises from the fact that the synchrotron radiation during
the studied flare is substantially diluted, presumably by radiation
produced at larger distances in a jet.

\subsection{Light-curves}
\begin{figure}
\resizebox{\hsize}{!}{\includegraphics{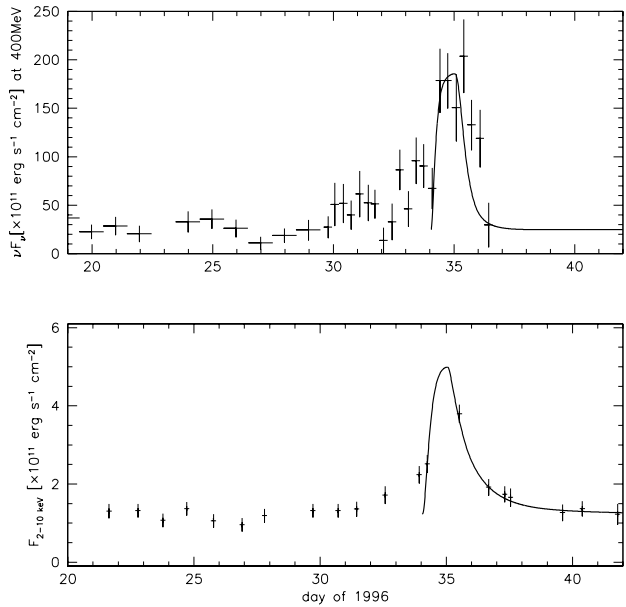}}
\caption{ 
  $\gamma$-ray (upper panel -- $\nu F_\nu$ flux at $400$ MeV)
  and X-ray (lower panel -- integrated $2$-$10$ keV flux) light-curves
  of the blazar 3C 279 during the Feb~'96 flare. Data points are from
  Wehrle et~al.~(\cite{weh98}). Thick, solid line shows the
  light-curves of our model.  }
\label{light}
\end{figure}
Fig.~(\ref{light}) presents the simulated light-curves together with
observational data points. Two curves are shown: one is the $\nu
F_\nu$ flux at $400$ MeV (EGRET range) and the second is the
integrated $2$-$20$ keV flux (XTE range). For both lightcurves the
steady components were added. These components were estimated from the
pre-flare observations. In the case of X-ray light curve the quiescent
emission is $F_{\rm q} = 1.23 \times 10^{-11}$erg s$^{-1}$ cm$^{-2}$
(Lawson et al.~\cite{lhm99}) and contributes up to $25$\% of the peak
flux. Thus for the X-ray band, where the radiation is presumably
dominated by the SSC process, the contribution of the quiescent
emission is significant. For gamma-rays we estimated the steady
component from data points before $30^{\rm th}$ day of 1996. Its value
is $\nu F_{\rm \nu,q} = 24.7 \times 10^{-11}$ erg s$^{-1}$ cm$^{-2}$
and its contribution to the peak flux is only $12$\%.

The agreement between the data and our model is remarkable taking into
account the relative simplicity of the model and complexity of the
real phenomena as suggested by, e.g., the gamma-ray light-curve.  The
discrepancy at the rising part of the X-ray light curve suggests an
existence of the soft X-ray precursors caused by Comptonization of the
external radiation by cold electrons (Sikora \& Madejski~\cite{sm02}).

\subsection{Spectral evolution}
\begin{figure}
\resizebox{\hsize}{!}{\includegraphics{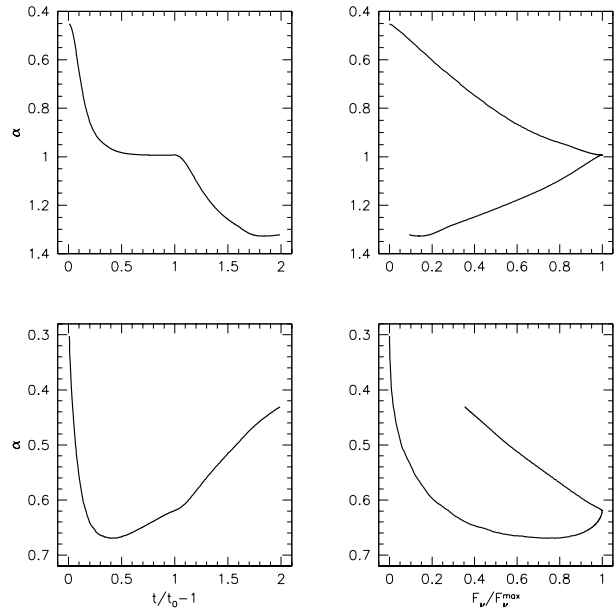}}
\caption{ 
  Spectral index $\alpha (F_\nu \propto \nu^{-\alpha})$
  evolution (left panels) and hardness-intensity diagram (right
  panels) of the blazar 3C 279 during the February 1996 flare. Upper
  panels show the index at $400$ MeV while lower panels show the index
  at $6$ keV. }
\label{specevol}
\end{figure}
In order to demonstrate spectral changes during the flare, in
Fig.~\ref{specevol} we present the evolution of the spectral index
$\alpha$ (defined by $F_\nu \propto \nu^{-\alpha}$) at two energies:
$400$ MeV and $6$ keV.  Two plots are presented for each energy: one
is the power law spectral index $\alpha$ as a function of time, the
second is the so called hardness--intensity diagram showing the
spectral index as a function of the flux.

Unfortunately, the comparison with observations in this case is
difficult.  There are no observations of the spectral index variation
during the flare by EGRET.  The reported slope of the $\gamma$-ray
spectrum for the event was $\alpha_\gamma = 0.97$ which is in
excellent agreement with the middle part of the curve in upper left
corner of the Fig.~\ref{specevol}.  For the XTE observations,
$\alpha_x$ ranges from $0.68$ to $1.26$, but the errors are large, and
there is a possible evidence of systematic error of $0.1$ (Lawson et
al.~\cite{lhm99}).  Even with the errors taken into account, the
observations seem to disagree with results presented in lower left
panel of the Fig.~\ref{specevol}.  However, we note that in this
range, the influence of the quiescent component is significant, and
the proper comparison would require the analysis which is beyond the
scope of this paper.  The steady component may also influence the
spectrum at $\gamma$-ray energies and cause the discrepancy for
observations outside of the maximum of the flare.

\subsection{Evolution of Electron Distribution}
\begin{figure}
\resizebox{\hsize}{!}{\includegraphics{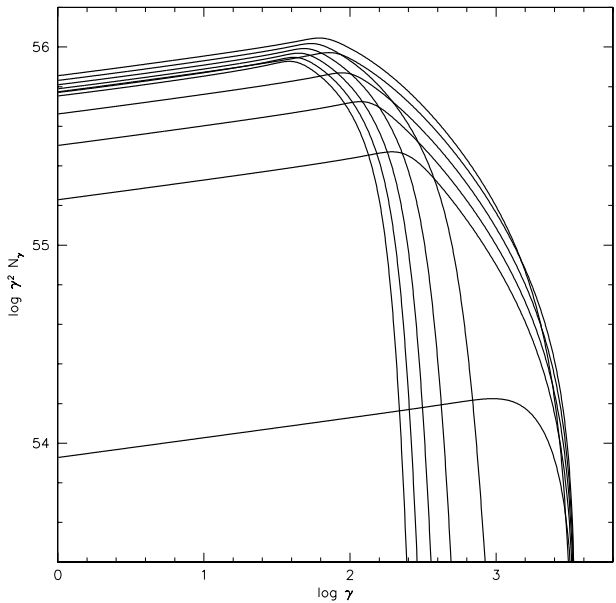}}
\caption{ 
  Evolution of the electron energy distribution during the
  February 1996 flare of the blazar 3C 279. Step between the curves is
  $\Delta r = 0.2 r_0$.  }
\label{eleevol}
\end{figure}
In Fig.~\ref{eleevol} we present the time evolution of energy
distribution of electrons during the flare.  The evolution is followed
from $r = r_0$ till the electrons reach the distance $3$ $r_0$.  At
the very beginning of the flare, the number of all electrons
increases, then the number of the most relativistic electrons starts
to saturate being balanced by the radiative energy losses. After the
injection of electrons stops, the high energy tail of the electron
distribution decays very rapidly, while only small changes can be
noticed at lowest energies.

The characteristic feature of this evolving electron spectrum is the
spectral break, which moves from higher to lower values as evolution
proceeds.  At a distance $r=2$ $r_0$, where injection stops and the
$\gamma^2 N_{\gamma}$ reaches maximum, the approximate location of the
break is $\gamma \simeq 65$.  The value calculated from Eq.~(\ref{gc})
is somewhat larger ($\gamma_c \simeq 105$) and this is because the
approximate analytical formula for $\gamma_{\rm c}$ does not include
synchrotron and SSC energy losses.  At energies lower than
$\gamma_{\rm c}$ (the slow cooling regime), the power-law energy
distribution of electrons $N_{\gamma} \sim \gamma^{-s}$ has an index
$s=p$, while for $\gamma \gg \gamma_{\rm c}$ (the fast cooling regime)
the slope of the electron distribution (due to efficient radiative
cooling) is steeper by unity, i.e., $s=p+1$.  The latter effect is
mostly obscured by the effect of the break at $\gamma_{\rm max}$.

\section{Summary}
In this article, we have presented our numerical code $BLAZAR$ which
simulates light-curves and spectra of blazars during flares.  In the
code, the structure of the source responsible for the production of
flares is approximated by thin uniform shells, propagating with
constant speed down the conical jet.  The shells are filled with
relativistic electrons/positrons, which are injected within a given
distance range at a constant rate.  Evolution of the electron energy
distribution is treated by the kinetic/continuity equation, given by
Eq.~(\ref{dNdtp}).  In that equation, the electron injection function
$Q$ is separated from the cooling term. This simplification can be
justified provided time scale of acceleration of electrons is much
shorter than time scale of their cooling and time scale of the shock
operation.  Since in FSRQ, which are main targets of the present
version of the code, the cooling break is located at $\nu_{\rm c} \ll
\nu_{\rm max}$, the above assumption is satisfied for all electrons
except for those with $\gamma \sim \gamma_{\rm max}$.  However, this
condition may not be satisfied for the X-ray selected BL Lac objects,
including TeV blazars. In these objects time scale of acceleration can
well be comparable to the lifetime of the source, and then
Eq.~(\ref{dNdtp}) must be modified in the way presented by Kirk et
al.~(\cite{kmp94}) (see also Kirk et al.~\cite{krm98} and B\"ottcher
\& Chiang~\cite{bch02} for application to blazars).

Very short gyration time scale (as compared with the radiative energy
losses) plus very likely presence of chaotic magnetic fields (because
of MHD turbulence induced around the shock fronts) justify the
assumption regarding the isotropic distribution of the electron
momenta, and therefore regarding the isotropy of the synchrotron and
SSC emission in the source co-moving frame. However, the isotropy
approximation doesn't apply to Comptonization of external
radiation. This is because due to relativistic aberration, the
external diffuse radiation is seen by the source as strongly beamed
from the front.  Because of this, the scattered radiation is
anisotropic - it is beamed into the source propagation direction
already even in the source co-moving frame.  This anisotropy,
originally pointed out by Dermer~(\cite{der95}), is self-consistently
treated in our code.  Radiative effects of the ERC anisotropy, often
ignored in other models, are particularly strong for $\psi_{\rm obs} >
1/\Gamma$.

Another advantage of our code is that it includes the adiabatic energy
losses in the kinetic equation for electrons.  In our model they are
related to the ``conical'' expansion of a shell.  Inclusion of
adiabatic losses is very important in calculating properly the
location of the cooling break $\nu_c$. Those losses also strongly
affect the light curves of flares observed at $\nu < \nu_{\rm c}$
(Sikora et al.~\cite{sbbm01}).

As an example of application of our code, we presented in \S 5 the
results of our modeling of the short term outburst observed in 3C 279
in February 1996 (Wehrle et al. 1998).  Our results demonstrate very
strong dilution of the synchrotron radiation component in this
object. This effect, often ignored completely by other models, can be
quite common, as suggested by much smaller amplitudes of optical
flares than of $\gamma$-ray flares (Ulrich et al.~\cite{umu97}). For
such objects, detailed fitting of the synchrotron spectrum in terms of
the homogeneous/one-component model is inappropriate and can lead to
very large errors in the model parameters. Our method, albeit
approximate, takes into account the dilution effect.

It is worth noting that the code ``BLAZAR'' was already used in
several previous papers. It was applied to the study the dependence of
amplitude of flares in blazars on the frequency, on the ratio
$r_0/\Delta r_{\rm coll}$, and on the time profile of an electron
injection function $Q(t')$ (Sikora et~al.~\cite{sbbm01}).  The code
was also used to study the relative role of hot dust and broad
emission line region as a source of the external seed photons for the
inverse Compton process in a jet (B{\l}a\.zejowski et al.) and was
applied to demonstrate the possible unification between the
MeV-blazars and GeV-blazars (Sikora et~al.~\cite{sbmm02}).  Finally,
after some small modifications allowing to model the radiation from
the shells propagating with a variable bulk Lorentz factor and
supplementing the code by relevant dynamical equations, it was used to
study the effect of lateral expansion of the collimated GRBs, possibly
resulting in the observed breaks in their light-curves (Moderski et
al.~\cite{msb00}).

In the present form, the code doesn't include the Klein-Nishina
effects nor the $\gamma$-ray absorption and e$^+$e$^-$ pair creation.
In FSRQ the Klein-Nishina effects become important at $\gamma > {\rm
a\ few} \times 10^4/\Gamma$, while the absorption -- at $h\nu \ge
20$GeV, both due to interactions with external BEL photons.  The high
energy break at $h\nu \sim 4-10$GeV, suggested by EGRET observations
of FSRQ during the $\gamma$-ray flares(Pohl et al.~\cite{pohl97}), and
the steep UV spectra seem to indicate that energy distribution of
electrons/positrons has a break or cutoff around $\gamma \sim {\rm a\
few} \times 10^4/\Gamma$ (see Appendix~A).  Hence, the KN and
$\gamma\gamma$ pair production effects are expected to be at most
marginal.  The above limitations, particularly the first one, are
expected to be relevant only for BL Lac objects, in particular for
those with strong TeV emission.  In the case of our model of the
3C~279 flare $\Gamma \gamma_{\rm max}h\nu_{\rm BEL}/ m_{\rm e} c^2
\simeq 0.5$ and $hv_{\rm max} = 20$GeV, so the effects are expected to
be negligible.

\begin{acknowledgements}
  This work was partially supported by the Polish Committee for
  Scientific Research grants no. 5 P03D 002 21 and PBZ 054/P03/2001.
  We are grateful for the hospitality of the Stanford Linear
  Accelerator Center, operated by Stanford Universty for the
  Department of Energy under contract no. DE-AC3-76SF00515, where some
  of the research described above was performed.  We would like also
  to thank Greg Madejski for his valuable comments which helped to
  improve the paper.
\end{acknowledgements}

\appendix
\section{Electron injection function}

In most FSRQ, the high energy spectra of flares can be recovered
assuming a single power-law injection function. Due to cooling effect,
the electron energy distribution evolves into the broken power-law,
which is reflected in the spectrum of the ERC component, with the
break located in the 1 - 30 MeV range (see \S 5). The low energy tails
observed in several FSRQ do not show any change of the slope down to
the lowest X-ray energies, and this seems to indicate that the
power-law electron injection function extends down to $\gamma_{\rm
min} \le $few (Tavecchio et al.~\cite{tav98}; Sikora et
al.~\cite{sbmm02}). This is independently supported by the detection
of circular polarization in radio cores of some quasars (Homan et
al.~\cite{haw01}), provided such polarization is generated by the
Faraday conversion process (Wardle et al.~\cite{war98}; Ruszkowski \&
Begelman~\cite{rb02}).

It is less clear how electron injection function behaves at the
highest energies and what is the maximum energy of injected electrons.
EGRET data of FSRQ during flares suggest existence of a high-energy
break at $4-10$ GeV (Pohl et al.~\cite{pohl97}). This high-energy
break cannot be caused by the Klein-Nishina effect, because in the
fast cooling regime, the luminosity of the Compton component is
determined by the electron injection function and reduction of the
Compton scattering cross-section in the Klein-Nishina regime can be
substantially compensated by the respective increase of the electron
densities. In particular, for electron injection function $\propto
\gamma^{-2}$ the Klein-Nishina portion of the Compton spectrum has
exactly the same spectral index, $\alpha=1$, as in the Thomson regime
(Zdziarski \& Krolik~\cite{zk93}).  The high-energy break also cannot
be produced by absorption of $\gamma$-rays by external photons. This
is because the radiative environment is transparent for photons
detected within the EGRET band (it becomes opaque only for $h\nu \ge
20$ GeV, due to absorption of such photons by optical/UV broad
emission lines). Hence, the break in $4-10$ GeV band, if real, must be
related to the break in the electron injection function, at $\gamma
\sim {\rm a\ few} \times 10^4/\Gamma$.

Is the injection function extending much above that break, but with a
steeper slope as suggested by observations of the so called
MeV-blazars (Sikora et al.~\cite{sbmm02}), or, there is a cutoff?  One
can try to answer that question by studying the spectra in the UV
band, where the high energy tails of the synchrotron component
dominate. However, the task is not easy, because synchrotron radiation
in this band can be affected by several factors: by extinction in the
host galaxy; by flattening of energy distribution of highest energy
electrons due to Compton scatterings in the KN regime (Dermer \&
Atoyan~\cite{da02}) and by $\gamma\gamma$ pair production; and, as in
the case of 3C~279, by dilution of the synchrotron component by
radiation produced at other locations in a jet.  Hence, the possible
extension of the injection function up to energies $\gamma \gg {\rm a\
few} \times 10^4/\Gamma$ can be verified observationally in FSRQ only
after the launch of the GLAST satellite in 2006.

Fortunately, the basic features of the current model of FSRQ as
adopted here are not very sensitive to the presence and details of
steep high energy tails of the injection function, and, in order to
reproduce the spectra at the level of detail required by the current
data, it is sufficient to use a single power-law injection function as
a first approximation.

\section{Broad emission line region}
\label{app2}

Reverberation mappings indicate that the BEL region in AGNs is
geometrically thick, with $r_{\rm max}/r_{\rm min} > 30$
(Peterson~\cite{pet93}).  The 3D structure of the BEL region is still
not fully understood, and in our code is assumed to be spherical. We
approximate the radial distribution of BEL luminosity by two power-law
functions which join at $r_{\rm BEL}$ where the luminosity has a peak,
i.e.
\bea 
\lefteqn{{\partial L_{\rm BEL} \over \partial \ln r} (r) = \null}
\nonumber \\ && = {\partial L_{\rm BEL} \over \partial \ln r}
(r_{\rm BEL}) \times \cases{ (r/r_{\rm BEL})^{q_1}& for $r < r_{\rm BEL}$ \cr
(r/r_{\rm BEL})^{-q_2} & for $r > r_{\rm BEL}$ \cr} \,
\eea
where from $L_{\rm BEL} = \int (\partial L_{\rm BEL}/\partial r)dr$
\be
{\partial L_{\rm BEL} \over \partial \ln r} (r_{\rm BEL}) = {q_1
q_2 \over q_1+q_2} L_{\rm BEL} \, .
\ee
and $q_1>0$ and $q_2>0$. In the case of a smooth transition from the
BEL region to the narrow emission lines region, one can expect $q_2
\sim 0.5$.  The value of $q_1$ is very uncertain and, in order to
minimize number of parameters, we adopt $q_1=q_2 = 0.5$.

\end{document}